\documentclass[aps,prr,twocolumn,superscriptaddress,preprintnumbers]{revtex4-2}
\usepackage{amsfonts, amssymb, amsmath, amsthm, mathrsfs}
\usepackage{qcircuit}
\usepackage{tikz}
\usepackage{braket}
\usepackage{enumitem}
\usepackage[hidelinks]{hyperref}
\hypersetup{
  colorlinks   = true, 
  urlcolor     = blue, 
  linkcolor    = blue, 
  citecolor   = blue 
}
\usepackage{subfigure}
\usepackage{graphicx}
\theoremstyle{definition}
\newtheorem{theorem}{\textit{Theorem}}
\newtheorem{lemma}{\textit{Lemma}}
\usepackage[ruled,vlined]{algorithm2e}
\tikzset{%
  every neuron/.style={
    circle,
    draw,
    minimum size=1cm
  },
  neuron missing/.style={
    draw=none, 
    scale=4,
    text height=0.333cm,
    execute at begin node=\color{black}$\vdots$
  },
}
\begin{document}
\title{Generative quantum learning of joint probability distribution functions}
\author{Elton Yechao Zhu}
\thanks{These authors contributed equally to this work.}
\affiliation{Fidelity Center for Applied Technology, \\
FMR LLC, Boston, Massachusetts 02210, USA}
\author{Sonika Johri}
\thanks{These authors contributed equally to this work.}
\affiliation{IonQ Inc, 4505 Campus Dr, College Park, Maryland 20740, USA}
\author{Dave Bacon}
\affiliation{IonQ Inc, 4505 Campus Dr, College Park, Maryland 20740, USA}
\author{Mert Esencan}
\affiliation{Fidelity Center for Applied Technology, \\
FMR LLC, Boston, Massachusetts 02210, USA}
\author{Jungsang Kim}
\affiliation{IonQ Inc, 4505 Campus Dr, College Park, Maryland 20740, USA}
\author{Mark Muir}
\affiliation{Fidelity Center for Applied Technology, \\
FMR LLC, Boston, Massachusetts 02210, USA}
\author{Nikhil Murgai}
\affiliation{Fidelity Center for Applied Technology, \\
FMR LLC, Boston, Massachusetts 02210, USA}
\author{Jason Nguyen}
\affiliation{IonQ Inc, 4505 Campus Dr, College Park, Maryland 20740, USA}
\author{Neal Pisenti}
\affiliation{IonQ Inc, 4505 Campus Dr, College Park, Maryland 20740, USA}
\author{Adam Schouela}
\affiliation{Fidelity Center for Applied Technology, \\
FMR LLC, Boston, Massachusetts 02210, USA}
\author{Ksenia Sosnova}
\affiliation{IonQ Inc, 4505 Campus Dr, College Park, Maryland 20740, USA}
\author{Ken Wright}
\affiliation{IonQ Inc, 4505 Campus Dr, College Park, Maryland 20740, USA}

\begin{abstract}
Modeling joint probability distributions is an important task in a wide variety of fields. One popular technique for this employs a family of multivariate distributions with uniform marginals called copulas. While the theory of modeling joint distributions via copulas is well understood, it gets practically challenging to accurately model real data with many variables. In this paper, we show that any copula can be naturally mapped to a multipartite maximally entangled state. Thus, the task of learning joint probability distributions becomes the task of learning maximally entangled states. We prove that a variational ansatz we christen as a `qopula' based on this insight leads to an exponential advantage over classical methods of learning some joint distributions. As an application, we train a quantum generative adversarial network (QGAN) and a quantum circuit born machine (QCBM) using this variational ansatz to generate samples from joint distributions of two variables in historical data from the stock market. We demonstrate our generative learning algorithms on trapped ion quantum computers from IonQ for up to eight qubits. Our experimental results show interesting findings such as the resilience against noise, outperformance against equivalent classical models and 20--1000 times less iterations required to converge as compared to equivalent classical models.\\

\end{abstract}

\maketitle
\section{Introduction}
Understanding the statistical relationship between several random variables is critical to all data-based analysis and decision-making. A few examples of its diverse applications include risk management \cite{James13}, portfolio optimization \cite{Markowitz52}, reliability analysis \cite{Zacks92}, recommender systems \cite{Aggarwal16}, climate research \cite{Storch99}, and medical imaging \cite{Palmer13}. Traditionally, single-parameter quantities such as the Pearson correlation or Spearman's correlation have been used to model dependence between variables. However, such measures are good for monotonic dependence, which is frequently too simplistic for real data. Data, such as that from the financial markets, engineering reliability studies, earth/atmospheric sciences tends to exhibit tail dependence. This means they do not appear to have much correlation but exhibit dependence in extreme deviations, as in the case of a black swan event \cite{Aloui11}.

Due to the reasons above, the relationship between random variables is now commonly modeled using a dependence function between uniformly distributed variables, called a ``copula". Sklar's theorem \cite{Sklar59}, which will be explained in the text, states that any multivariate joint distribution can be written in terms of univariate marginal distributions and a copula that describes the dependence structure between the variables. This provides the theoretical foundations for the use of copulas. Since then, copulas have found many applications in quantitative finance \cite{vandenGoorbergh05}, engineering \cite{Kilgore11}, and medicine \cite{Lahorgue17}.

Some of the commonly used copulas include elliptical and Archimedean. However, empirical copulas (copulas from real data) tend to be a mixture of copulas and are commonly modeled using parametric methods like maximum likelihood estimation \cite{Joe97}. As a rule, the more complex the copula and the more completely it describes the data, the more computationally challenging it becomes to extend it to higher dimensional data.

More recently, generative models have been proposed for statistical modeling. These learn to generate data with the same statistics as a given training dataset, effectively learning its distribution. The model can be used to output new samples that could plausibly have belonged to the original dataset \cite{Goodfellow14,Diederik14}. One of the most successful of these is the generative adversarial network (GAN). In GAN, two neural networks compete with one another in a minimax game. In Ref. \cite{Goodfellow14}, the authors showed theoretically that a GAN will learn the data distribution if given enough capacity, data and training time. Since then, GANs have found applications in areas like data augmentation \cite{Antoniou17}, fashion design \cite{Wu16}, and super resolution \cite{Ledig17}, etc. However, there is still debate over whether GANs will successfully learn a given distribution \cite{Arora18}. In particular, problems like vanishing or unstable gradient, mode collapse and nonconvergence, sensitivity to hyperparameters, and either the generator or discriminator overpowering the other, can make them challenging to train \cite{GANissues}.

In this context, the question arises whether quantum computers can provide any advantage over classical models for generating samples that reproduce the inter-dependence between multiple variables of a given dataset. Quantum wavefunctions naturally represent probability distributions which can be hard to sample from by classical algorithms. Quantum computers can also be used to generate correlations which cannot be efficiently reproduced by classical means \cite{Aaronson11,Boixo18}. Therefore, intuitively, learning dependence between multiple random variables is something at which a quantum computer may excel. Further, an application that requires sampling from a probability distribution is likely one of the near term applications of quantum computers. As evidence, note that this is already the basis of the quantum supremacy demonstration \cite{Arute19}.

Quantum generative models such as the quantum circuit born machine (QCBM) \cite{Song18,Benedetti19} and quantum generative adversarial network (QGAN) \cite{Dallaire18,Lloyd18} have been proposed as quantum algorithms for the learning of both classical and quantum data. Experimental demonstrations include training a QCBM to produce sample from the bars and stripes dataset \cite{ZhuBAS} and to approximately prepare many-body states \cite{Benedetti19}. QGANs have been used in state preparation \cite{Zoufal19}, learning of qubit channels \cite{Hu19}, and image generation \cite{Huang20}. In Ref. \cite{Rudolph20}, the output from a quantum circuit is used as the prior to a classical GAN. Since quantum algorithms are known to achieve speedup over classical algorithms in a variety of tasks \cite{Grover96,Shor97}, it has been hypothesized that quantum generative models can do better than classical generative models in terms of expressivity, stability during training, and shorter training time \cite{Lloyd18}. 

In the existing literature, generative quantum learning has either been restricted to learning quantum data, univariate distributions, or multivariate distributions without separating the correlation structure from the marginal distributions. Here, we extend this prior work using a natural connection between entanglement and the copula function. We proposed an architecture based on sampling of variational circuits. This allowed us to argue quantum advantage based on shallow circuits. This also extends previous quantum advantage argument in QGAN based on fault-tolerant quantum algorithms like the factoring algorithm\cite{Shor97,Dallaire18} or the quantum algorithm for systems of linear equations \cite{Harrow09,Lloyd18}. As the model uses shallow circuits to learn classical data, it is NISQ-friendly and of wide applicability. Another significant difference to previous QGAN architectures is that we consider measurement samples as outputs, which saves a lot of computational resources when compared to other schemes that uses expectation values as outputs. Our tests on quantum simulators and hardware show advantages for the stability of the training process as well which is relevant for practical applications.

This paper is especially relevant as quantum hardware has also been advancing rapidly implying that useful applications of quantum algorithmic advances may not be far off. A variety of noisy intermediate scale quantum (NISQ) computers are now available over the cloud, of which we use the trapped ion quantum processing units (QPU) from IonQ.

The outline of our paper is as follows. We begin by introducing the basic concepts of GAN, QGAN and QCBM necessary for this work in Sec. \ref{sec:QGANQCBM}. In Sec. \ref{sec:copulaentanglement}, we introduce copulas and show how they can be learned using maximally entangled quantum states. Sec. \ref{sec:hardware} gives a brief overview of the quantum hardware used for our experiments. In Secs. \ref{sec:QGAN} and \ref{sec:QCBM}, we devise training routines for our quantum generative models that are implementable on a NISQ device. We then demonstrate via numerical simulation and executions on the IonQ QPUs that our generative learning algorithms can learn from empirical data. Sec. \ref{sec:evaluation} compares the performance of our algorithms with a classical GAN of similar number of parameters as well as classical parametric methods, revealing advantages for the quantum methods. We argue via communication and computational complexity that our quantum learning algorithms have a computational advantage over classical algorithms for learning joint distributions in Sec. \ref{sec:advantage}. Sec. \ref{sec:future} provides an overview of future work. We present a discussion and conclusion in Sec. \ref{sec:conclusion}.

\section{Classical and Quantum Generative Adversarial Network, Quantum Circuit Born Machine} \label{sec:QGANQCBM}
Generative modeling is an unsupervised learning task that involves the learning of patterns in the training data such that the model can generate new examples as if they are drawn from the original data. A generative adversarial network is a framework to train a generative model by transforming it to a supervised learning task, in which two neural networks are trained alternatively in a minimax game.

GAN employs two sub--models, a generator $G$ and a discriminator $D$. $G_{\theta_g}: Z\to X$ represents a mapping from a latent space $Z$ to the data space $X$, with $\theta_g$ denoting the parameters of $G$. $D_{\theta_d}: X\to [0,1]$ represents a mapping from the data space $X$ to a scalar, with $\theta_d$ denoting the parameters of $D$. For each $x\in X$, $D(x)$ can be interpreted as the probability that $x$ belongs to the original training data. $D$ is trained to maximize the probability of differentiating true data from samples generated by $G$. Simultaneously, $G$ is trained to ``fool" the discriminator by generating realistic data samples. In other words, the two models $D$ and $G$ are trained in a two-player minimax game with value function
\begin{align}\label{eqn:loss}
    \min_G&\max_D V(D,G)\nonumber\\
         &=\mathbb{E}[\log D(x)]+\mathbb{E}[\log \left(1-D(G(z))\right)].
\end{align}
After training, the generator is retained to generate new data samples for downstream use.

In practice, the minimax optimization is realized with two loss functions, one for the generator and one for the discriminator. For a batch of $m$ samples from the real data $\{x^{(l)}\}$ and $m$ noise vectors $\{z^{(l)}\}$, the loss function of the generator is
\begin{equation}
    L_G(\theta_g,\theta_d)=-\frac{1}{m}\sum_{l=1}^{l=m}[\log D(G(z^{(l)}))]
\end{equation}
and the loss function of the discriminator is
\begin{equation}
    L_D(\theta_g,\theta_d)=-\frac{1}{2m}\sum_{l=1}^{l=m}[\log D(x^{(l)})+\log (1-D(G(z^{(l)})))].
\end{equation}

A typical GAN training schedule is show in Algorithm \ref{alg:GAN}.

\begin{algorithm}
\SetAlgoLined
\KwResult{Trained Generator $G$}
 Initialize Network\;
\For{$i\leftarrow 1$ \KwTo $iterations$}{
  Sample minibatch of $m$ noise vectors $\{z^{(l)}\}$ \;
  Sample minibatch of $m$ data examples $\{x^{(l)}\}$ \;
  Calculate Discriminator Loss $L_D$ \;
  Update the discriminator by descending its stochastic gradient $\nabla_{\theta_d}L_D(\theta_g,\theta_d)$ \;
  Calculate Generator Loss $L_G$ \;
  Update the generator by descending its stochastic gradient $\nabla_{\theta_g}L_G(\theta_g,\theta_d)$ \;
 }
 \caption{GAN Training Loop}
 \label{alg:GAN}
\end{algorithm}

The stochastic gradient descent is usually performed using adaptive algorithms like Adam\cite{Kingma14} with a suitable learning rate. Some researchers also update the discriminator/generator multiple times in a training iteration or use different learning rates, depending on the problem.

Intuitively, the generator is trying to get better at generating data samples and thus fool the discriminator. The discriminator is trying not to be fooled. The discriminator provides positive feedback to the generator, such that the generator continues to improve. However, since it is a minimax optimization, it is very common for GAN training to suffer from non-convergence (GAN fails to learn well from the training data) and mode collapse (GAN generates data with limited variety). While many techniques have been proposed to improve the stability of GAN training, these are still common problems researchers face.

Recently, quantum generative adversarial network (QGAN) \cite{Dallaire18,Lloyd18} has been proposed as a generalization of classical GAN that can run on a quantum computer. QGAN can be used to learn either classical or quantum data. For the learning of quantum data, each training sample $x$ becomes a quantum state $\ket{x}\in \mathcal{H}_X$. $G_{\theta_g}: \mathcal{H}_Z\to \mathcal{H}_X$ represents a quantum circuit from the latent Hilbert space $\mathcal{H}_Z$ to the data space $\mathcal{H}_X$, and $D_{\theta_d}:\mathcal{H}_X\to [0,1]$ can be either a quantum circuit or a classical neural network. Note that if $D$ is a classical neural network, initial measurement is needed to convert $\ket{x}$ to some classical data. 

For the learning of classical data, one way of realizing the quantum generator $G_{\theta_g}:Z\to X$ is to encode a noise vector $z$ as some quantum state $\ket{z}$, apply a quantum circuit $U_G$, and then measure the expectation of some observable $O_G$, i.e., $G(z)=\bra{z}U_G^\dag O_G U_G\ket{z}$. Another way, as noted in Ref. \cite{Zoufal19}, is to represent $G_{\theta_g}$ as a quantum circuit $U_G$ followed by some POVM (positive operator-valued measure) $M_x$ in $X$ space. In this case, the dependence of $G$ on a latent space $Z$ is optional as POVM measurement is inherently probabilistic and data samples are drawn from the probability distribution
\begin{align}
    q(x)=\text{tr}(M_xU_G\ket{0}\bra{0}U_G^\dag).
\end{align}

A quantum circuit born machine (QCBM) is a quantum generative model similar to the generator part of QGAN, except the training objective is different. In the QCBM case, if we suppose that the target distribution is $p$, a QCBM seeks to minimize the distance between distributions $p$ and $q$ directly, through either Kullback-Leibler (KL) divergence \cite{Benedetti19} or maximum mean discrepancy \cite{Wang18}. Here, we use the KL divergence as a cost function, defined as 
\begin{align}
    d_{\text{KL}}=\sum_{x} p(x) \log\frac{p(x)}{q(x)}.
\end{align}

\section{Copula and Quantum Entanglement}\label{sec:copulaentanglement}
Given random variables $(\mathcal{X}_1,\ldots,\mathcal{X}_d)$, probability integral transform states that the marginal cumulative distribution functions defined as $F_i(x)=\text{Pr}[\mathcal{X}_i\leq x]$ satisfies $\text{Pr}[F_i(\mathcal{X}_i)\leq u]=u$. Therefore the random variables
\begin{equation}\label{eqn:transform}
    (\mathcal{U}_1,\ldots,\mathcal{U}_d)=(F_1(\mathcal{X}_1),\ldots,F_d(\mathcal{X}_d))
\end{equation}
have marginals, which are uniformly distributed on $[0,1]$.

A copula $C$ of $(\mathcal{X}_1,\ldots,\mathcal{X}_d)$ is defined as the joint cumulative distribution function of $(\mathcal{U}_1,\ldots,\mathcal{U}_d)$:
\begin{equation}\label{eqn:copula}
    C(u_1,\ldots,u_d)=\text{Pr}[\mathcal{U}_1\leq u_1,\ldots,\mathcal{U}_d\leq u_d].
\end{equation}

Reversing the above steps gives a way to generate samples from multivariate distributions. Given a procedure to generate samples $(\mathcal{U}_1,\ldots,\mathcal{U}_d)$ from the copula function $C$,
\begin{equation}\label{eqn:retransform}
    (\mathcal{X}_1,\ldots,\mathcal{X}_d)=(F_1^{-1}(\mathcal{U}_1),\ldots,F_d^{-1}(\mathcal{U}_d))
\end{equation}
would be samples from the original multivariate distribution.

Sklar's theorem \cite{Sklar59,Durante13} states that every multivariate cumulative distribution function on random variables $(\mathcal{X}_1,\ldots,\mathcal{X}_d)$ can be expressed in terms of their marginal cumulative distribution functions $F$ (defined as $F_i(x)=\text{Pr}[\mathcal{X}_i\leq x]$) and a copula $C$ [defined in Eq. (\ref{eqn:copula})], i.e.,
\begin{align}
    \text{Pr}[\mathcal{X}_1\leq x_1,\ldots,\mathcal{X}_d\leq x_d]=C(F_1(x_1),\ldots,F_d(x_d)).
\end{align}

Copulas are used to simulate correlated variables because they remove the structure of the marginal distributions, and capture just the point-wise correlation between variables. Simple copula formulas are extensively used in finance, engineering and medicine. However these simple formulas often fall short in accurately capturing the relationship between variables. Instead more sophisticated or empirical copulae can be used but these become hard to model and sample from as the number of variables increases. Very recently, generative models have been successfully used to model copulas \cite{Patki16}.

For our study, we test our generative models using the following framework:
\begin{enumerate}[label={(\arabic*)}]
    \item Use Eq. (\ref{eqn:transform}) to transform training data into copula space
    \item Fit known copula model/GAN/QGAN/QCBM in copula space
    \item Sample from known copula model/GAN/QGAN/QCBM in copula space
    \item Use Eq. (\ref{eqn:retransform}) to transform synthetic data from copula space back to original space
\end{enumerate}

Next, we show that every copula with density can be represented by a maximally entangled state.

Suppose $C:[0,1]^d\to [0,1]$ is a $d$-dimensional copula with density $c:[0,1]^d\to [0,1]$. $c$ can be obtained as
\begin{align}
    c(u_1,\ldots,u_d)=\frac{\partial^d C(u_1,\cdots,u_d)}{\partial u_1\cdots \partial u_d}.
\end{align}
This can be represented as a quantum state
\begin{align}
    \ket{c}=&\sum_{k_1,\ldots,k_d=0}^1\int_{[0,1]^d}e^{i\phi(u_1,\cdots,u_d,k_1,\cdots,k_d)}\frac{\sqrt{c(u_1,\cdots,u_d)}}{2^{d/2}}\nonumber\\
    &\times\ket{u_1,k_1}\cdots\ket{u_d,k_d} d^d u.
\end{align}
Since $\mathcal{U}_1,\ldots,\mathcal{U}_d$ have uniform marginals,
\begin{align}
    \int_0^1 \dots \int_0^1 c(u_1,\cdots,u_d) du_1\cdots du_{i-1}du_{i+1}\cdots du_d=1 ~~~\forall i,
\end{align}
and by setting the phases $\phi$ appropriately, the reduced density matrix $\rho_i=\text{Tr}_{1,\cdots,i-1,i+1,\cdots,d}(\ket{c}\bra{c})$
for partition $i$ is completely mixed and $\ket{c}$ is a maximally entangled state with respect to each of the $d$ partitions.(See Appendix \ref{appd:copulaproof} for the detailed proof.)

\begin{figure*}[htb]
\centering
\begin{subfigure}(a)
\includegraphics[width=0.6\columnwidth]{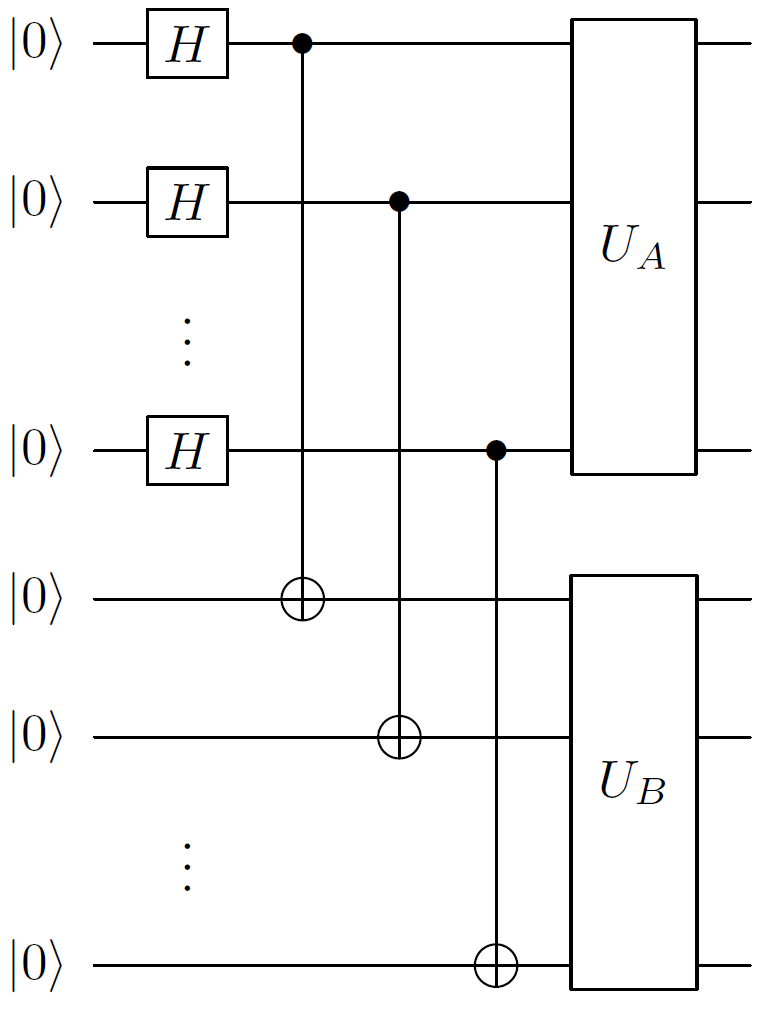}
\end{subfigure}
\centering
\hspace{2em}
\begin{subfigure}(b)
\includegraphics[width=1\columnwidth]{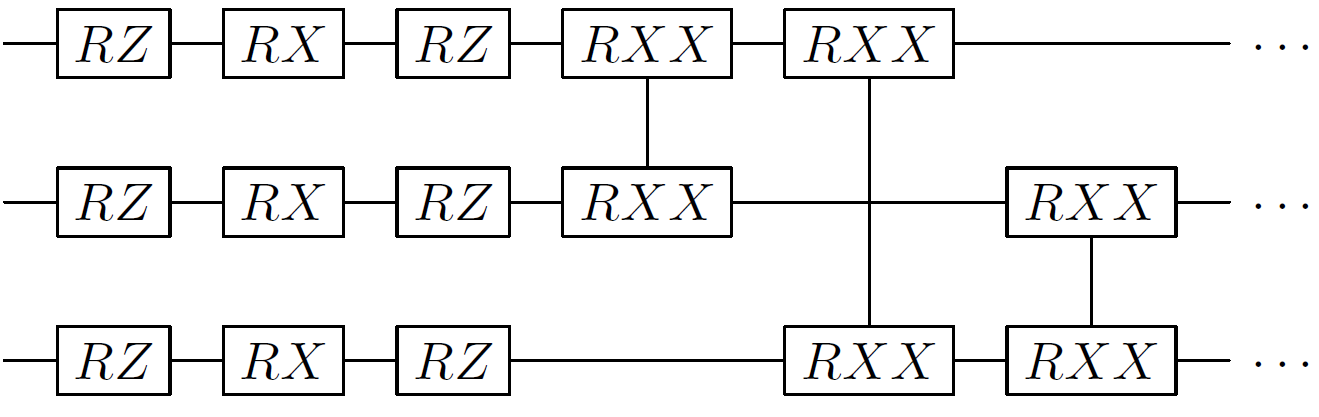}
\end{subfigure}
\caption{(a) The qopula circuit for 2 random variables. The top (bottom) half of the circuit consists of qubits that belong to the register, which provides samples of the first (second) random variable. (b) The ansatz for $U$ corresponding to 3 qubits. All gates are parametrized by angles which are optimized during the learning process. The structure can be repeated for multiple layers each with different parameters. Here the 2-qubit RXX gates represent $\exp(i\theta X_i X_j)$ acting on qubits $i$ and $j$, which can be executed as $\text{CNOT}(i,j)\exp(i\theta X_i)\text{CNOT}(i,j)$ using standard gates available on gate-model quantum computers. The RZ and RX gates represent rotations around the Z and X axes.}
\label{fig:copula}
\end{figure*}

Here we construct a quantum circuit that can prepare such maximally entangled states, which we christen as a qopula circuit. The quantum circuit for two random variables is shown in Fig. \ref{fig:copula}(a). 
The first part of the circuit consists of Hadamard gates applied to all qubits in register A, followed by CNOTs between A and B. This creates $N_q/2$ Bell pairs, which are distributed between A and B. This results in the creation of a state of the form
\begin{align}
    \ket{\psi_{AB}}=\frac{1}{2^{N_q/4}}\sum_{i=0}^{2^{N_q/2}-1} \ket{i_A}\ket{i_B}.
\end{align}

In this state, the reduced density matrix of the states in each register is completely mixed ($\rho_A=I/|A|$). In the next step, unitaries act locally on registers A and B. Since $U^\dagger I U=I$, the action of the unitaries does not change the reduced density matrix, but it does change the correlations between the registers A and B. Note that the unitaries on each register can be set by parameters that are independent of the other.

If each register is measured in the computational basis with output $x_1,\ldots,x_n,y_1,\ldots,y_n$, the measured bitstrings can represent discretized values of each random variable in the domain $[0,1]$ through the following linear transformation
\begin{align}\label{eqn:lineartransform}
    x=\frac{1}{2^{1+n}}+\sum_{i=0}^n\frac{1}{2^{n}}x_n.
\end{align}
Therefore this ansatz can model the correlations between uniform random variables and capture the copula dependence structure.

The number of qubits in each register determines the discretization of the probability distribution. The more the qubits, the more finely the point-wise correlation can be learnt. If the number of qubits is restricted, for practical application, one can use the available qubits to learn the significant bits and then append random bits in the position of the less significant bits.

The circuit can be extended to more than two variables by preparing GHZ states instead of Bell pairs as the starting point on which the operators $U$ act. The exact structure of $U$ is not restricted and a subject of future research. Here we use a structure that consists of layers alternating between a ``driver" layer that consists of parametrized single qubit rotations RZ and RX and an ``entangler" layer that consists of $RXX$ gates as shown in Fig. \ref{fig:copula}(b) \cite{Maslov17}. This structure can be repeated many times to approximate general unitaries. We also note that the uniform marginal condition can be satisfied by many states  other than the simple Bell or GHZ states \cite{Enr_quez_2016}. 

\begin{figure*}[t]
    \subfigure[]{
        \includegraphics[width=0.65\columnwidth]{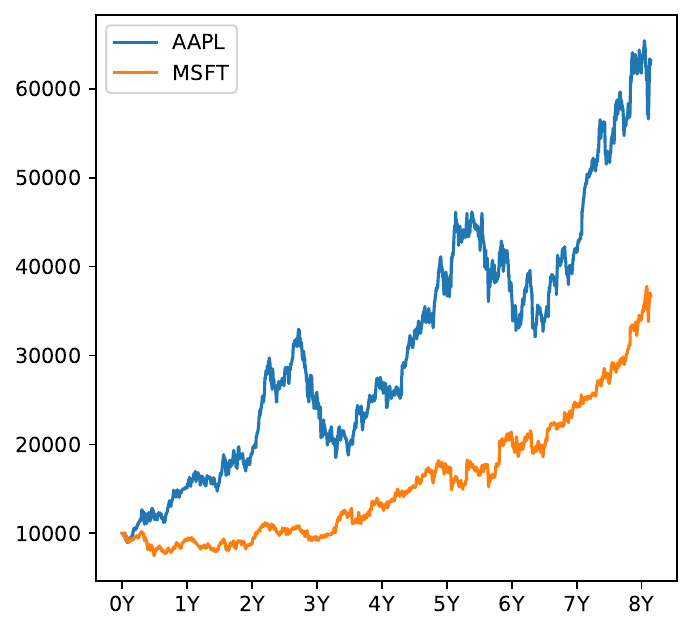}
    }
    \subfigure[]{
        \includegraphics[width=0.65\columnwidth]{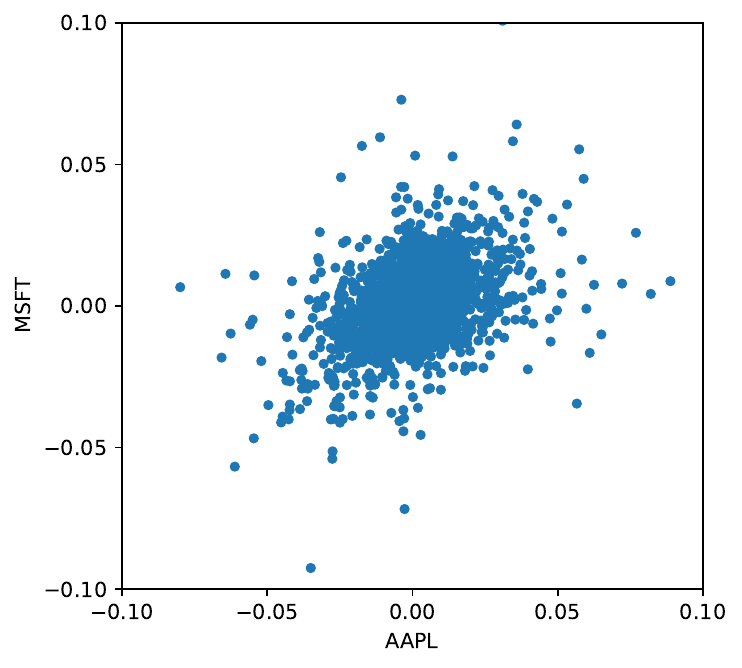}
    }
    \subfigure[]{
        \includegraphics[width=0.65\columnwidth]{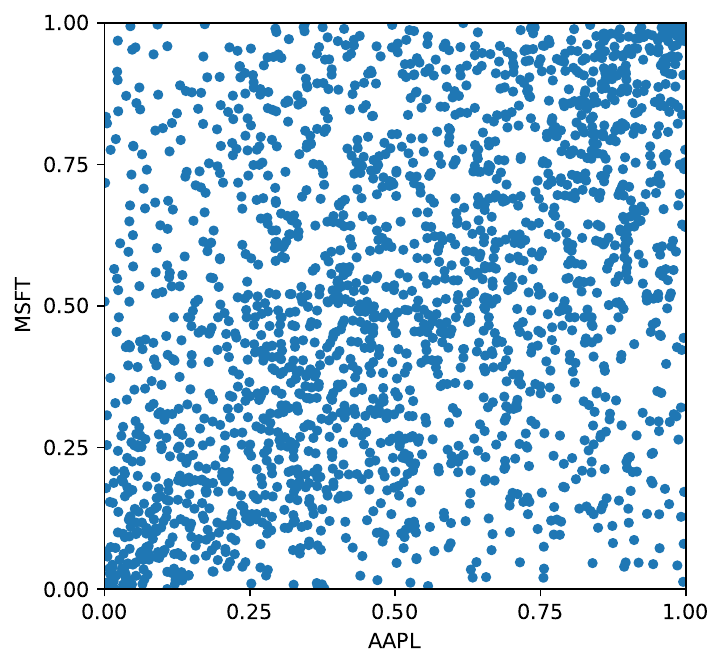}
    }
    \caption{(a) Hypothetical growth of \$10 000 invested in AAPL and MSFT between 2010--2018. (b) Scatter plot of daily returns. (c) Scatter plot of data after probability integral transform.}
    \label{fig:data_visualization}
\end{figure*}

The dataset that we use is the daily return of AAPL and MSFT between 2010--2018. Figure \ref{fig:data_visualization}(a) represents the hypothetical growth of capital if \$10 000 is invested in both stocks (assuming dividends are reinvested). The daily returns are calculated as the percentage change of close prices between adjacent trading days. Suppose on Day 1 a stock's close price is $P_1$ and its close price is $P_2$ on Day 2, then its daily return on Day 2 is $r=(P_2-P_1)/P_1$. The daily return data is plotted both in $\mathcal{X}$ space [real data space, Fig. \ref{fig:data_visualization}(b)] and $\mathcal{U}$ space [copula space, Fig. \ref{fig:data_visualization}(c)]. In model training, no time series model is assumed and hence the daily returns are assumed to be independent events.

While the dataset being used for this research comes from the financial market, we would like to emphasize that our quantum generative algorithms are completely general and should be able to learn from any dataset that has correlations among different fields.

\section{Quantum Hardware}\label{sec:hardware}

The experiments were run on IonQ trapped ion quantum processing units (QPUs), which utilize stable ground states of trapped ytterbium atomic ions as qubits. Two ground states of these atoms with hyperfine splitting are used as the two qubit states, and they are initialized, manipulated and detected with adequately tailored laser lights applied to them. The quantum circuit is executed by applying quantum logic gates driven by Raman transitions using up to 32 individually addressed beams of 355 nm light. These Raman transitions are programmed to provide a universal gate set by applying individual rotations of a given qubit’s internal states to realize a single-qubit gate, as well as two-qubit gates by inducing Molmer-Sorenson type interaction to generate entanglement between a pair of qubits in the system. IonQ’s QPUs feature all-to-all connectivity, where a two-qubit gate can be applied directly to any pair of qubits in the system regardless of their physical location \cite{Wright19}. IonQ’s QPUs available on the cloud are self-calibrating, which means that the system monitors the quality of the quantum logic gates constantly and makes sure they are fully calibrated before the computational tasks are executed. The hardware used in this work utilizes IonQ QPUs that are available over the cloud, as well as a next-generation QPU.

\section{Quantum Learning through QGAN}\label{sec:QGAN}
In our QGAN implementation, the quantum generator comprises of 6 qubits, and each unitary uses 1 layer of ansatz as described in Fig. \ref{fig:copula}. Therefore the quantum generator has 24 trainable parameters. A measurement in the computational basis is performed at the end of the circuit and Eq. (\ref{eqn:lineartransform}) is applied to get the 2-dimensional sample in copula space. To deal with the issue of discretization, the measurement output of $ (x_0,x_1,x_2,y_1,y_2,y_3) $ is extended to $(x_0,x_1,\cdots,x_{22},$ $y_0,y_1,\cdots,y_{22})$ with the additional 40 bits uniformly randomly generated, and then the linear transformation is applied.

The discriminator is a feed-forward classical neural network with input dimension 2, a hidden layer of dimension 32 and an output layer of dimension 1. The input and hidden layer comprises of a linear layer and leaky rectified linear unit (ReLU), whereas the output layer is a linear layer followed by sigmoid function.

To compare the expressivity of QGAN vs GAN, we design the generator of classical GAN to have the same number of trainable parameters. This can be achieved by a 2-layer feed-forward neural network with input dimension 6 and output dimension 2. The input layer consists of a linear layer followed by batch norm and ReLU. The output layer consists of a linear layer followed by a sigmoid function. A sigmoid function is needed to transform samples back to the unit square.

Neural-network models are typically trained using gradient-based optimizers. In the case where the cost function has a quantum circuit component, the gradient with respect to circuit parameters can be calculated using the parameter-shift rule \cite{Schuld19}. However, this procedure requires two executions of the original circuit (each with shifted parameters), and thus the number of circuit executions scales two times as the number of circuit parameters. For example, in our case, it would take 48 circuit executions to compute the gradient with respect to the 24 parameters in the quantum generator. This is executed sequentially in most software for quantum computers. In addition to slower evaluation of the cost function itself, since commercially available quantum computers (including IonQ's) are typically cloud based, sequential evaluation will also be slow due to network latency. For this reason, we avoided the computation of quantum gradients and used the simultaneous perturbation stochastic approximation (SPSA) algorithm \cite{Spall87}. SPSA algorithm updates the parameter in an iterative process, with the $k$--th step 
\begin{align}
    \theta_{k+1}=\theta_k-a_k\hat{g}_k(\theta_k),
\end{align}
where the $i$--th component of the gradient estimator $\hat{g}_k(\theta_k)$ is
\begin{align}
    \left(\hat{g}_k(\theta_k)\right)_i=\frac{C(\theta_k+c_k\Delta_k)-C(\theta_k-c_k\Delta_k)}{2c_k(\Delta_k)_i}.
\end{align}
Here $\Delta_k$ is a random perturbation vector with each element drawn uniformly at random from $\{-1,1\}$. For initial learning rate $a$ and step size $c$, $a_n$, and $c_n$ are updated as $a_k=a/k$ and $c_k=c/k^\gamma$, where $\gamma$ is chosen from $[1/6,1/2]$. Through numerical simulation, we observe that running SPSA algorithm with 3--5 iterations in place of a gradient descent step for the generator, yields good performance in our QGAN training schedule. Since each iteration step only requires 2 circuit executions, an optimization step using SPSA would only require 6--10 circuit executions in total.

Therefore, our QGAN training schedule will be as shown in Algorithm \ref{alg:QGAN}.
\begin{algorithm}
\SetAlgoLined
\KwResult{Trained Generator $G$}
 Initialize Network and angles of the Quantum Circuit ansatz\;
 \For{$i\leftarrow 1$ \KwTo $iterations$}{
  Run the quantum generator with $m$ shots to generate $m$ measurements \;
  Sample minibatch of $m$ data examples $\{x^{(l)}\}$ \;
  Calculate Discriminator Loss $L_D$ \;
  Update the discriminator by descending its stochastic gradient $\nabla_{\theta_d}L_D(\theta_g,\theta_d)$ \;
  Calculate Generator Loss $L_G$ \;
  Update the Quantum Generator by using SPSA algorithm \;
 }
 
 \caption{QGAN Training Loop}
 \label{alg:QGAN}
\end{algorithm}

We first test our model in simulation before running it on IonQ's cloud QPUs. All simulations and experiments of QGAN are trained with 1000 iterations and use the parameter $a=0.008$, $c=0.01$, number of iterations $n_{iter}=5$, $\gamma=0.101$ for SPSA, learning rate 0.0015 for the discriminator, batch size $m=2048$. To reduce the experiment time, we used the maximum learning rates possible without breaking down the learning. To better understand the effect of hardware noise, random initialization with the same seed is used for comparing simulation and experiment results. However, other random initializations were also used in simulation and results are summarized in Sec. \ref{sec:evaluation}.

\begin{figure*}[t]
    \subfigure[]{
        \includegraphics[width=1\columnwidth]{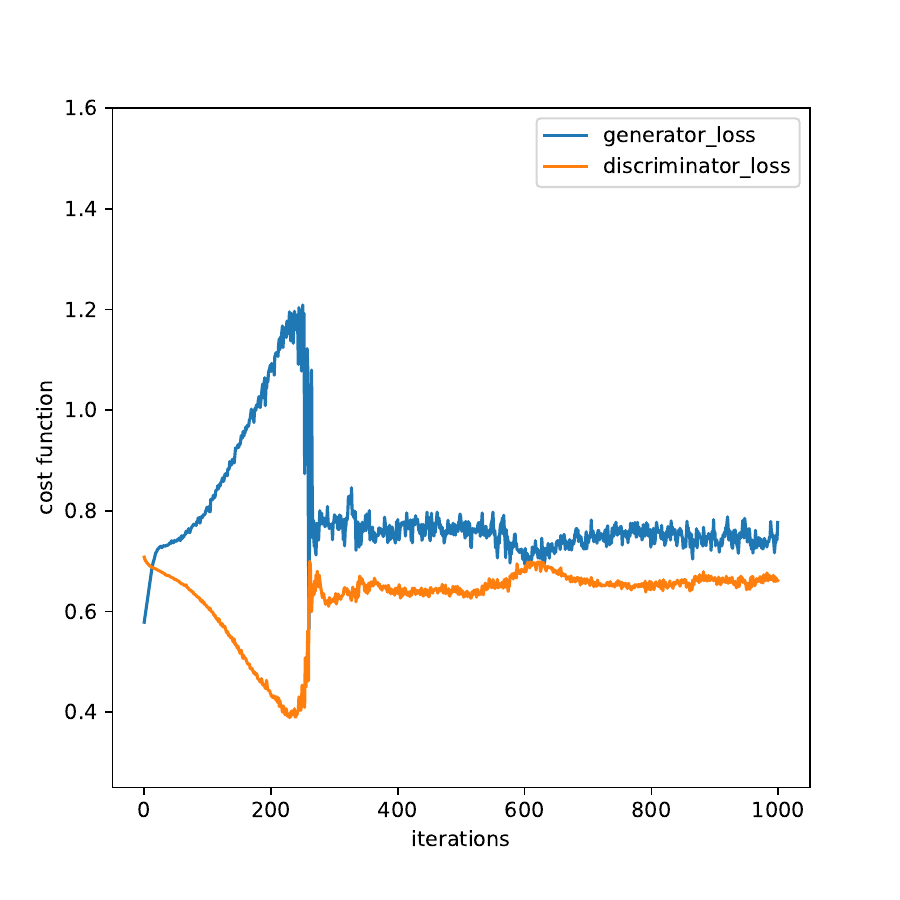}
    }
    \subfigure[]{
        \includegraphics[width=1\columnwidth]{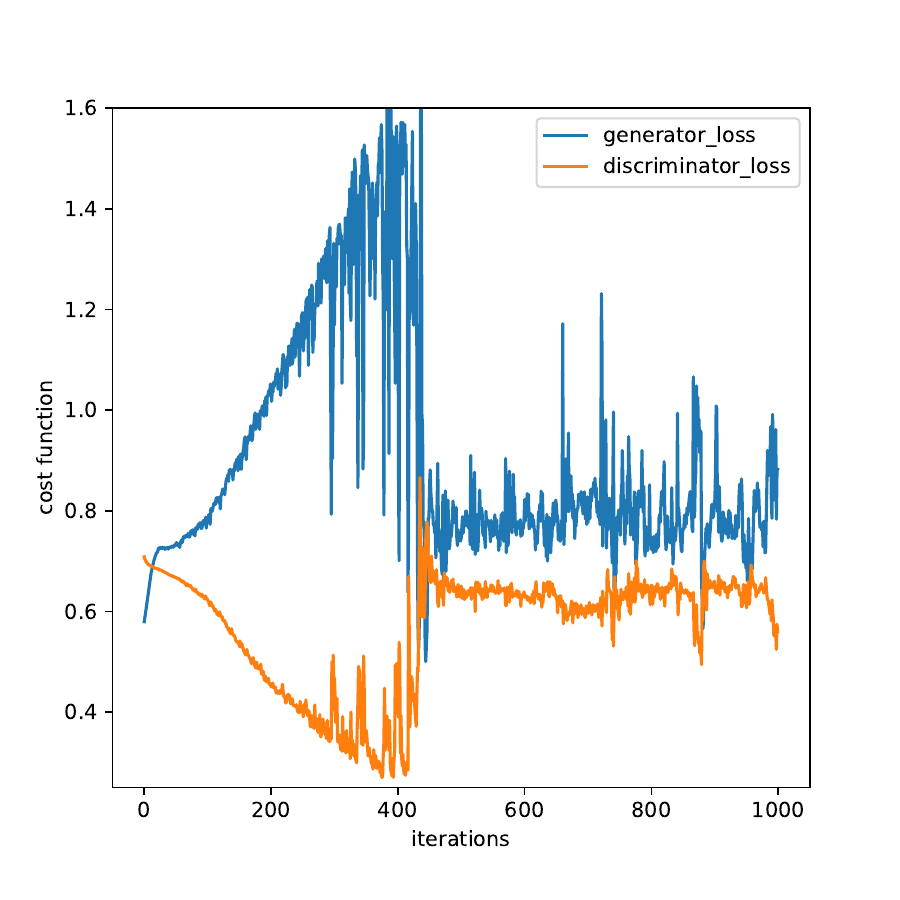}
    }
    \caption{Plot of the loss functions from QGAN training for $N_q=6$ qubits. (a) Loss plot from simulator. (b) Loss plot from experiment.}
    \label{fig:loss_qgan}
\end{figure*}

\begin{figure*}[t]
    \subfigure[]{
        \includegraphics[width=1\columnwidth]{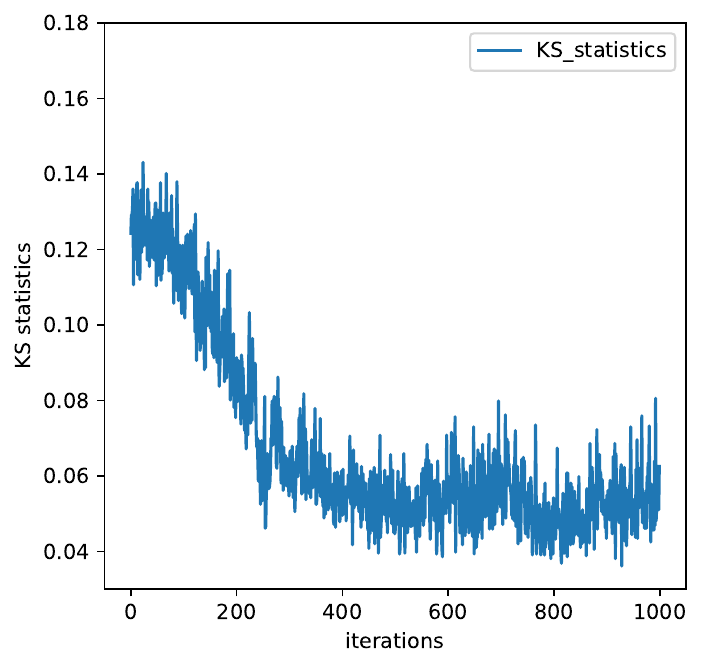}
    }
    \subfigure[]{
        \includegraphics[width=1\columnwidth]{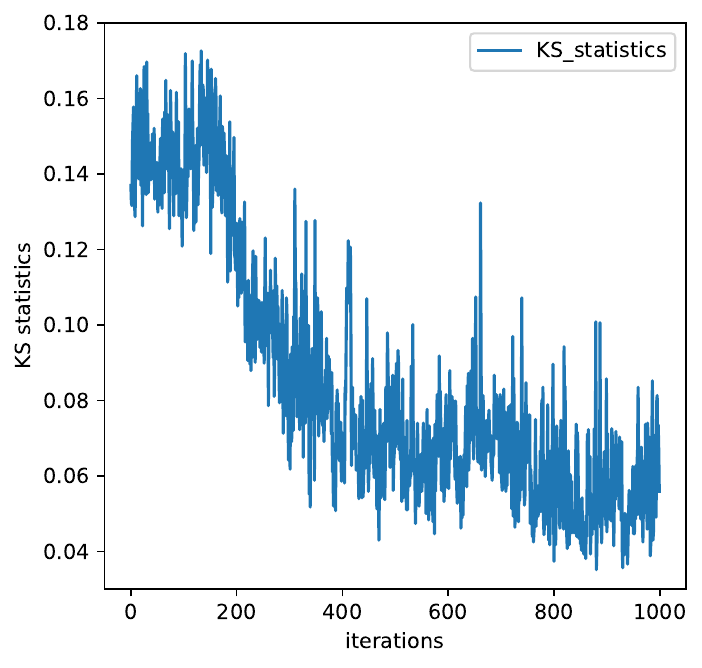}
    }
    \caption{Plot of the KS statistics from QGAN training for $N_q=6$ qubits. (a) KS statistics plot from simulator. (b) KS statistics plot from experiment.}
    \label{fig:ks_plot}
\end{figure*}

Results from the simulator, Fig. \ref{fig:loss_qgan}(a) indicates that in our training process, the losses from the generator and discriminator first diverge before converging around iteration 250. This implies that, the discriminator learns faster than the generator in the beginning, but then the generator experiences an Aha! moment and quickly catches up. This is also evident when we examine the quality of synthetic data produced (Fig. \ref{fig:ks_plot}), where the KS statistics rapidly decreases around the convergence point, and then enters a stable phase. (KS statistics is explained more in Sec. \ref{sec:evaluation} where we evaluate the performance of the models.)

As quantum computers at the moment still have significant noise, it is essential to investigate how noise would affect our training algorithm. We ran the same training algorithm on a mixed state simulator with the addition of up to $4\%$ depolarizing noise following each two-qubit gate, which is an appropriate model for noise on IonQ's cloud QPUs \cite{Johri20}. We observe that the losses of the generator and discriminator diverge a bit further. This is to be expected that noise makes the generator a weaker learner and therefore it needs more time to learn. However, in most of the cases, our QGAN converges eventually (anywhere between iterations 300--500), indicating robustness against noise and suitability to run on NISQ machines. We observe that in our experiment running on IonQ's cloud QPUs, the losses converged at around iteration 400, agreeing with results from noisy simulations. 

\begin{figure*}[t]
    \subfigure[]{
        \includegraphics[width=1\columnwidth]{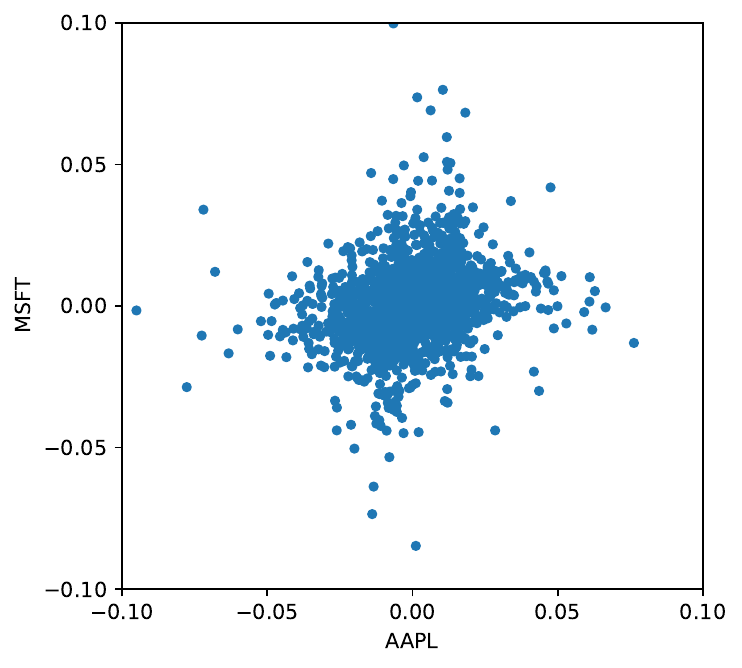}
    }
    \subfigure[]{
        \includegraphics[width=1\columnwidth]{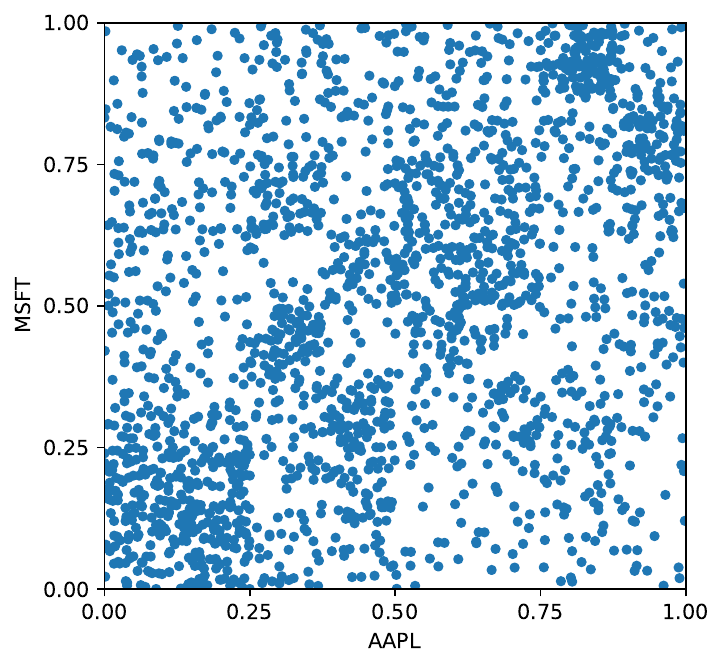}
    }
    \caption{(a) Scatter plot of QGAN-generated daily returns after reverse probability integral transform. (b) Scatter plot of QGAN-generated copula data. $N_q=6$ qubits were used.}
    \label{fig:synthetic_data_visaulization}
\end{figure*}

QGAN-generated data is visualized in both $\mathcal{X}$ space [real data space, Fig. \ref{fig:synthetic_data_visaulization}(a)] and $\mathcal{U}$ space [copula space, Fig. \ref{fig:synthetic_data_visaulization}(b)]. We observe that the data in copula space [Fig. \ref{fig:synthetic_data_visaulization}(b)] are concentrated around a few clusters and are less evenly distributed than the original data [Fig. \ref{fig:data_visualization}(c)]. This is because we used only 3 qubits to represent each dimension and therefore the output is very discrete. One would only see 64 unique data points had we not added the random bits after measurements. The random bits have the effect of creating small fluctuations around the original discrete value, hence the clusters.

\section{Quantum Learning through QCBM}\label{sec:QCBM}
For the QCBM, we use the KL divergence as a cost function. As the target, we bin the data in copula space into $2^{N_q/2}$ bins, where $N_q$ is the number of available qubits, generating a discrete probability distribution $p$. This is compared to  the probability distribution $q$ over the computational basis states measured at the output of the circuit. The KL divergence cost function is 
\begin{align}
    d_{\text{KL}}=\sum_{i,j} q_{i,j} \log\frac{q_{i,j}}{p_{i,j}},
\end{align}
where $i$ and $j$ are for the first and second variable respectively. Numerically, we implement the clipped version of the KL divergence where $p_{i,j}$ are set to $10^{-6}$ if below that value. The QCBM training schedule is shown in Algorithm \ref{alg:QCBM}.

\begin{algorithm}
\SetAlgoLined
\KwResult{Trained Quantum Circuit}
 Initialize angles of the Quantum Circuit ansatz\;
 \For{$i\leftarrow 1$ \KwTo $iterations$}{
  Run the quantum circuit with $m$ shots to generate $m$ measurements \;
  Calculate the KL divergence $d_{KL}$ \;
  Update the circuit angles by using SPSA algorithm \;
 }
 
 \caption{QCBM Training Loop}
 \label{alg:QCBM}
\end{algorithm}

\begin{figure*}[t]
    \begin{subfigure}[]{}
        \includegraphics[width=1\columnwidth]{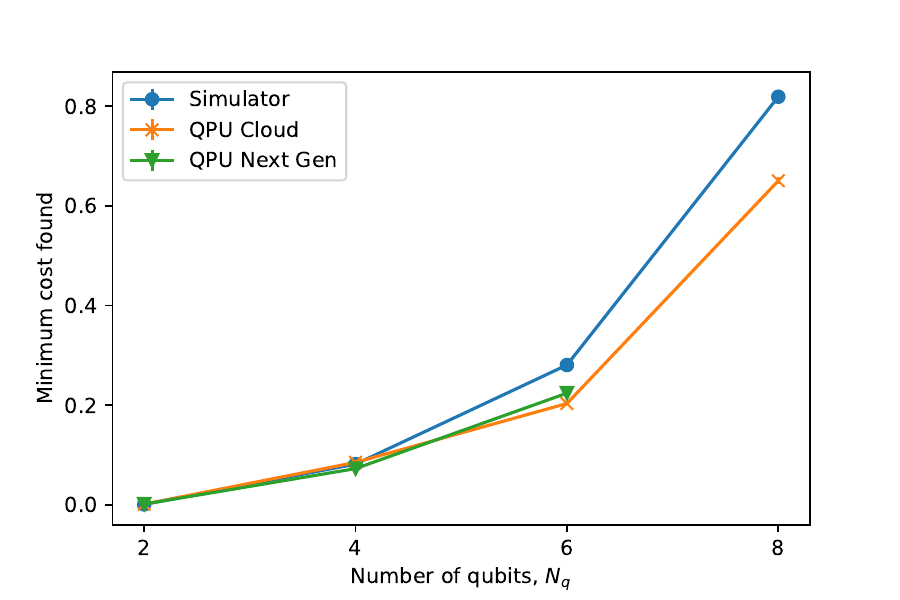}
    \end{subfigure}
    \begin{subfigure}[]{}
         \includegraphics[width=1\columnwidth]{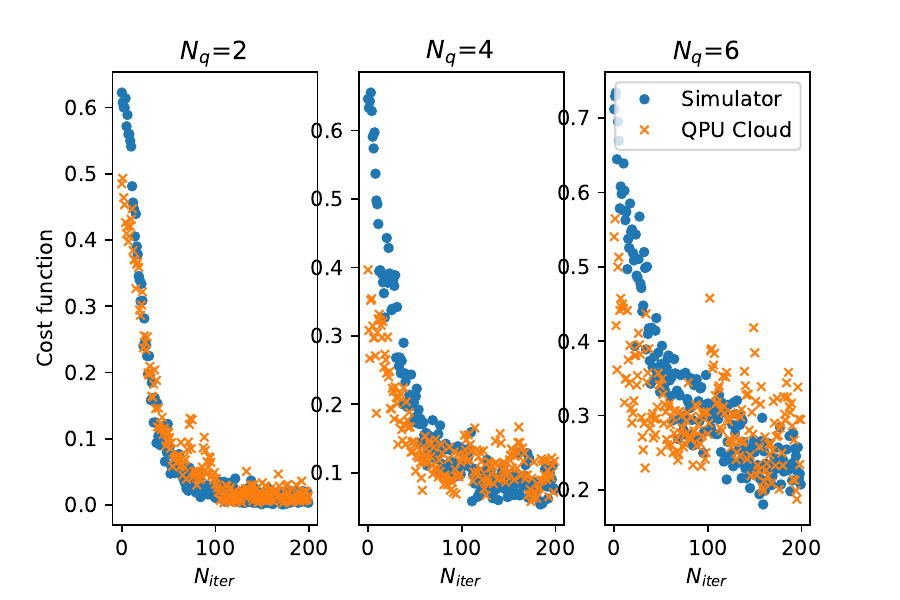}
    \end{subfigure}
    \caption{QCBM training (a) average of the 10 minimum values of the cost function reached during training as a function of the number of qubits for optimization parameters $a=0.5$, $c=0.5$. The error bars are smaller than the marker sizes. (b) Convergence of the cost function as a function of the number of iterations for $a=0.1$, $c=0.1$. The panels indicate results from different numbers of qubits $N_q$.}
    \label{fig:cost_qcbm}
\end{figure*}

For converting the measured bits into samples, Eq. (\ref{eqn:lineartransform}) is used along with the procedure of appending the random bits as described in the QGAN section.

The results of training are shown in Fig. \ref{fig:cost_qcbm}. The results are shown for a series of system sizes consisting of 2, 4, and 6 qubits. A one-layer ansatz was used in each case. Each system size was initialized with the optimal angles from the previous smaller system size, which can be thought of as a form of transfer learning. The training was done on the simulator, IonQ's cloud QPUs, and its latest generation QPU, which is not yet accessible through the cloud.

The optimization algorithm used for training is the SPSA similar to the QGAN training. Two hundred iterations were used for all experiments. Figure \ref{fig:cost_qcbm}(a) shows results for up to 8 qubits on the simulator and cloud quantum computer and up to 6 qubits on the next-generation quantum computer. The optimization parameters used were $a=0.5$ and $c=0.5$, for which the training finds the minimum within 20 iterations. As can be seen in this figure, which shows the average over the 20 smallest values of the measured cost function, for 2 and 4 qubits, all three backends converge to a similar value. For 6 qubits, somewhat surprisingly, we observe that the noisy hardware has a smaller value of the final cost function. This observation of the noise being favorable for converging to a low value of the cost function at the beginning of the optimization is further confirmed in Fig. \ref{fig:cost_qcbm}(b), which shows the convergence for $a=0.1$ and $c=0.1$, where it is slow and hence can be seen clearly. We note that we do not expect that the noise will continue to remain beneficial at larger system sizes, however.

While vanishing gradient (barren plateau) has been observed in several variational quantum algorithms \cite{McClean18}, we note that we see no evidence of it even as the system size increases. Instead, the training landscape becomes noisier even for the simulator as shown by the broadening of the cost function values as the number of qubits increases. This is to be expected since only a finite number of shots are used whereas the number of possible outputs increases exponentially with the number of qubits. We note that this would also be true for a classical generator. More discussion of the QCBM training can be found in Appendix \ref{appd:QCBMtraining}.

We further note that for the cases we tested, the QCBM technique requires fewer function calls to train the model than the QGAN approach. Which technique may end up performing better will depend on the details of the application. Finally, we also note that an improved convergence is not observed for the increased quantum gate fidelity of the next generation QPU for either approach, implying that the noise plays some positive role in the statistical machine learning models. The major advantage of the next generation QPU in this application is its increased stability that leads to reduced runtime of the entire optimization task by about a factor of 2.

\section{Evaluation of Model Performance}\label{sec:evaluation}
We use the 2-dimensional 2-sample Kolmogorov-Smirnov (KS) test \cite{Peacock83} to evaluate our generative models together with classical GAN and a parametric model. The null hypothesis of the 2-sample KS test is that the two samples are drawn from the same distribution, and the alternative hypothesis is that they are drawn from different distributions. It is typically performed on 1-dimensional distributions but here we are performing on 2-dimensional distributions. We use a significance level of 0.05 here. If the p-value produced by the test is more than the significance level specified, then we accept the null hypothesis. In the case where multiple models produce samples which are accepted by the KS test, we use the associated KS statistics to compare among the models. The KS statistic quantifies the distance between the empirical distributions of two samples. For 1-dimensional data, it can be defined as
\begin{align}
    D_{KS}(P,Q)=\sup_x|P(X\leq x)-Q(X\leq x)|,
\end{align}
where $P$ and $Q$ are the empirical cumulative distributions. Please refer to Ref. \cite{Peacock83} for the 2-dimensional generalization.

For the classical models and QGAN, 2048 samples are collected from our generative models, and for the QCBM, 2000 samples are collected from 4 circuits with minimal cost function that were run with 500 shots each. These are compared with a bootstrapped version of the training data. The architecture of classical GAN is described earlier. The parametric model we used involves fitting the copula data with a Gaussian copula.
\begin{table*}[t]
    \caption{KS statistics and p-value of KS test across multiple models. The quantum models use $N_q=6$ qubits.}
    \begin{ruledtabular}
    \begin{tabular}{lcccc}
         Model & $D_{KS}$ (the smaller the better) & p-value (threshold 0.05)\\
         \hline
         Parametric model&0.0449&0.117\\
         Classical GAN&0.0363--0.0508&0.0530--0.309\\
         QGAN simulation& 0.0320--0.0396&0.226--0.473\\
         QGAN experiment, QPU cloud &0.0352&0.3570\\
         QCBM simulation&0.0425--0.0520&0.0511--0.1717\\
         QCBM experiment, QPU cloud&0.0373--0.0515& 0.0548--0.3030\\
         QCBM experiment, QPU Next Gen&0.0330--0.0510&0.0578--0.4465\\
    \end{tabular}
    \end{ruledtabular}

    \label{tab:modelperformance}
\end{table*}

\begin{figure}[h]
\centering
        \includegraphics[width=1\columnwidth]{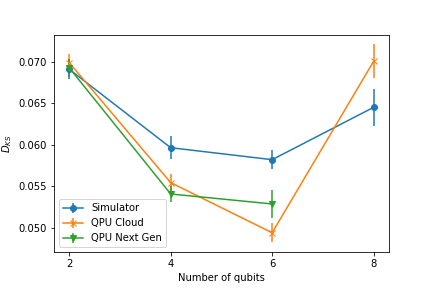}
    \caption{KS metric averaged over the 10 best converged values of the cost function as a function of system size for the QCBM.}
    \label{fig:ks_qcbm}
\end{figure}

For classical GAN and QGAN on simulator, we were able to run the model multiple times with different initializations and thus there is a range of results. For QGAN, we initialized the parameters uniformly at random from $[0,2\pi]$ and note that our model converges regardless of initialization. For classical GAN, we used the default initialization method from PyTorch \cite{pytorch19} with different seeds. We note that about $40\%$ of our model instances are accepted by the 2D KS test with threshold 0.05. The rest failed to learn and were rejected by the test. In Table \ref{tab:modelperformance}, the metrics of those accepted instances were included. For the QCBM, in order to see the range of performance, we first generated 5 datasets, each of which consisted of data from running the QCBM with parameters from 4 unique points where the cost function values were minimized during the optimization. For each dataset, we calculated the performance by affixing less significant random bits to the ones from the generator. We then also averaged over 20 such sets of random bits. The range shown in Table \ref{tab:modelperformance} is then the minimum and maximum of the $D_{KS}$ and p-values of these 5 datasets compared to the target distribution.

As one can see, the parametric model does a good job in modeling the copula but generative models are able to do better. The result from QGAN is consistent and outperforms classical GAN with similar number of parameters. The QCBM performs slightly worse than the QGAN but comparable to the classical GAN. Figure \ref{fig:ks_qcbm} shows the results for the $D_{KS}$ as a function of number of qubits for the QCBM training. Up to 6 qubits, the quantum computers perform better for QCBM training than the simulator as is to be expected from their better performance in copula space as well. We note that $D_{KS}$ decreases up to 6 qubits as expected but increases after that indicating the need for more shots and iterations for the optimization to converge.

We note that it is entirely possible for classical GAN to achieve better results with a deeper neural network as generator. However, the same can be said about our quantum generative models as they are only learning the significant bits at the moment. With better hardware, more qubits, more layers and better design for the ansatz $U$, our quantum generative models will also do better. Nevertheless, our results provide compelling evidence that our quantum generator has more expressivity than classical generative models.

We also note that we are able to train QGAN/QCBM at a much faster learning rate and therefore conclude the training with much fewer iterations than classical GAN. In classical GAN, the learning rate used is 0.0001 and model training concludes after 20000 iterations. Attempts to increase the learning rate failed due to non-convergence in model training. In QGAN, model training concludes after 1000 iterations. In QCBM for 6 qubits, the training converges to a good value for as little as 20 iterations.

Currently the gate speeds of commercially available QPUs are typically slower than those of classical computers. Additionally, due to the current access mode of IonQ QPUs over the cloud that includes queued access priority, network latency and calibration time on the machine, our QGAN experiment took 2 weeks to complete. At this time, this is much slower as compared to other models. The training time for our classical GAN is 6 minutes on a CPU-only machine and that of QGAN simulation is 4 minutes. For the QCBM on the next-generation quantum hardware, the training for 6 qubits with 200 iterations took approximately 5 hours whereas it took 9 hours on the cloud system. We expect many of the network latencies and calibration procedures to improve on QPUs in the future, as well as dedicated access to be available. Further, as shown in the next section, there will be certain joint probability distributions for which the run time for the classical models and the simulations are expected to grow exponentially with the number of qubits, while the QPU time will only grow polynomially, ultimately making the classical and quantum technologies competitive for learning applications.

\section{Quantum Computational Advantage}\label{sec:advantage}
We next present arguments for the presence of computational advantage in our quantum machine learning algorithms over classical machine learning algorithms such as neural networks.

The first source of advantage stems from the quantum supremacy argument of ``instantaneous quantum polynomial" (IQP) circuits \cite{Bremner11}, which is a family of non-universal quantum circuits. The output distribution of IQP circuits is of the form
\begin{align}\label{eqn:IQPdistri}
    q^{IQP}(z)=|\bra{z}H^{\otimes n}U_DH^{\otimes n}\ket{0^n}|^2
\end{align}
where $U_D=\exp(iD)$ and
\begin{align}
    D=\sum_{k,l}J_{k,l}Z_kZ_l+\sum_kM_kZ_k.
\end{align}
Here we quote some results from Ref. \cite{Bremner11}.
\begin{lemma}[Corollary 1 of Ref. \cite{Bremner11}]
If the output probability distributions generated by uniform families if IQP circuits could be weakly classically simulated to within multiplicative error $1\leq c \leq \sqrt{2}$ then the polynomial hierarchy would collapse to its third level, i.e., $PH=\Delta_3$\cite{ph}.
\end{lemma}
Hence, this type of distributions cannot be efficiently sampled classically, assuming $PH=\Delta_3$. The same argument underlies quantum supremacy arguments of boson sampling \cite{Aaronson11} and sampling outputs of QAOA \cite{Farhi16}.

Here we briefly sketch the original arguments. We encourage interested readers to refer to Ref. \cite{Bremner11} for detailed proofs. A quantum circuit with postselection has output registers $O$ and post-selection registers $P$. Instead of sampling measurement results $x$ directly from measurement on the output registers, we consider only those runs of the process for which a measurement on the postselection registers $P$ yield $0\dots0$. In this construction, we require the circuit to have the property $\text{Pr}[P=0\dots0]\neq 0$ so that the conditional distribution $\text{Pr}[O=x|P=0\dots 0]$ is well defined.  If A and B are complexity classes, $A^B$ denotes the class A with an oracle for B \cite{Arora09}.

While IQP $\subsetneq$ BQP (BQP = bounded error quantum polynomial time), Ref. \cite{Bremner11} argued that with post-selection, they have the same power, i.e., post-IQP = post-BQP. If there is an efficient classical algorithm that can sample outputs from IQP, then IQP $\subseteq$ BPP (BPP = bounded error probabilistic polynomial time). The same relation would hold with postselection, i.e., post-IQP$\subseteq$ post-BPP. This would imply post-IQP = post-BQP $\subseteq $ post-BPP. Since we know a universal quantum computer can simulate a classical computer efficiently, i.e., BPP $\subseteq$ BQP, the same would hole with post-selection, i.e., post-BPP $\subseteq$ post-BQP. This would immediately imply post-BQP = post-BPP. The equality of these two complexity classes will imply $PH=\Delta_3$, which we believe will not happen. Therefore, there is no efficient classical algorithm that can sample outputs from IQP, assuming the polynomial hierarchy does not collapse to the third level.
Here we present a similar result for qopula circuits.
\begin{theorem}
If the output probability distributions generated by uniform families if qopula circuits could be weakly classically simulated to within multiplicative error $1\leq c \leq \sqrt{2}$ then the polynomial hierarchy would collapse to its third level, i.e., $PH=\Delta_3$.
\end{theorem}
\begin{proof}

We note that the output distribution of IQP circuits is a special case of the conditional output distribution from our circuit. The output distribution of our circuit can be written as
\begin{align}
    q(z_A,z_B)=\left|\bra{z_A,z_B}U_A\otimes U_B\left(\frac{\ket{0_A0_B}+\ket{1_A1_B}}{\sqrt{2}}\right)^{\otimes N_q/2}\right|^2.
\end{align}
If we set all the parameters in $U_B$ to 0 such that $U_B$ is trivial, then
\begin{align}\label{eqn:condistri}
    q\left(z_A|z_B=0\cdots 0\right)=|\bra{z_A}U_A\ket{0^{\otimes N_q/2}}|^2.
\end{align}

If we set $U_A$ to use 1 layer of the ansatz and set angles of all the RZ rotations to 0, then we are left with $RX$ and $RXX$ gates. Since they commute, we can write $U_A=\exp(i\tilde{D})$ where $\tilde{D}=\sum_{k,l}J_{k,l}X_kX_l+\sum_kM_kX_k.$ We can see that Eq. (\ref{eqn:condistri}) is of the same form as Eq. (\ref{eqn:IQPdistri}), as $HZH=X$. Since post-selecting the qopula circuit with $B=0\dots 0$ allows us to generate distributions identical to that of IQP circuits, we see that IQP $\subseteq$ post-Qopula (the complexity class of qopula circuits with postselection) and therefore post-IQP $\subseteq$ post-Qopula $\subseteq $ post-BQP. Since  post-IQP = post-BQP \cite{Bremner11}, we get post-Qopula = post-BQP. If the output distribution of qopula circuits can be sampled efficiently classically, we would have Qopula $\subseteq $ BPP and therefore post-Qopula $\subseteq$ post-BPP. This would imply post-BQP $\subseteq $ post-BPP, which would result in $PH=\Delta_3$.
\end{proof}

Reference \cite{Coyle20} argued quantum advantage of QCBM based on IQP circuits.

The second source of advantage stems from the exponential separation between shallow quantum circuits and shallow classical circuits \cite{Watts19}. Reference \cite{Watts19} defines a problem called the Parity Halving Problem PHP$_n$ as outputing a string $ y\in \{0,1\}^n $ given an input $ x\in \{0,1\}^n $ of even parity, such that $ |y|\equiv |x|/2 ~(\text{mod}~2) $. Here we present a result in Ref. \cite{Watts19} that is relevant for our discussion. We encourage interested readers to refer to Ref. \cite{Watts19} for detailed proofs.
\begin{lemma}[Theorem 2 in Ref. \cite{Watts19}]\label{lem:PHP}
The Parity Halving Problem ($PHP_n$) can be solved exactly by a $QNC^0/\ket{CAT}$ circuit. But on the uniform distributions over all valid inputs (even parity strings), any $AC^0/rpoly$ circuit of depth $d$ and size at most $\exp{(n^{1/{10d}})}$ only solves the problem with probability $\frac{1}{2}+\exp{(-n^\alpha)}$ for some $\alpha>0$.
\end{lemma}

Here $\ket{CAT}$ denotes the GHZ-type state $\frac{1}{2}(\ket{0^n}+\ket{1^n})$. AC$^0$/rpoly is the family of Boolean circuits of depth $O(1)$ and polynomial size, with unlimited-fanin gates, with the ability to sample from any probability distribution on polynomially many bits that is independent of the input, but that can depend on the input size. The quantum circuit that solves the problem is given as $H^{\otimes n}RZ(\vec{x}\pi/2)$ acting on $\ket{CAT}$, followed by measurement in the standard basis. Here $RZ(\vec{x}\pi/2)$ is understood as a layer of $RZ$ gates where each gate is $RZ(x_i\pi/2)$. The measurement outcome is interpreted as the $y\in \{0,1\}^n$ required by the problem.

Based on the above results, we derive the following.
\begin{theorem}
There exists $n$ sets of measurements (each of size 2) determined by the optimizer (as defined by the corresponding machine learning model) such that a classical circuit of depth $d$ with randomized advice of size poly(n) with size at most $\exp{(n^{1/{10d}})}$ can only produce samples from the output distribution produced by qopula circuits with probability $\frac{1}{2}+\exp{(-n^\alpha)}$ for some $\alpha>0$.
\end{theorem}
\begin{proof}
The quantum circuits proposed in the proof of Lemma \ref{lem:PHP} are special cases of qopula circuits, because
\begin{equation}
 H=iRZ(\pi/4)RX(\pi/4)RZ(\pi/4)   
\end{equation}
and thus the circuit is equivalent to qopula circuit for $n$ random variables, with 1 qubit for each random variable and 1 layer. The angle parameters are $\pi/4$ for the $RZ$'s followed by $\pi/4$ for the $RX$'s and $\pi/4+\vec{x}\pi/2$ for the $RZ$'s. There is no two-qubit gate after the initial entangling procedure.

The quantum generative models described in this paper involves parameterized quantum circuits and an optimizer producing the corresponding measurements (SPSA based on KL divergence for the case of QCBM and SPSA based on binary cross entropy and the discriminator for the case of QGAN). Replicating the proof of Lemma. \ref{lem:PHP}, it would mean that, in the case the parameters provided by the optimizer comes from the uniform distribution of even parity binary strings, the quantum circuits would produce a quantum state that is the uniform superposition of the halved-parity binary strings. If the corresponding quantum circuit is replaced by a classical circuit of depth $d$ with randomized advice of size polynomial in the input size, such classical circuits with size at most $\exp(n^{1/10d})$ can only produce samples from the output distribution produced by qopula circuits with probability $\frac{1}{2}+\exp(-n^\alpha)$ for some $\alpha>0$. If the classical circuit is in the form of a classical neural network, it would mean the neural network with constant depth would need exponential dimension (either the input or the hidden dimension) to simulate the output distribution exactly.
\end{proof}
\renewcommand\qedsymbol{$\blacksquare$}

The third source of advantage stems from Bell's theorem. The discussion below in summarized in Appendix \ref{appd:bell} in the form of a theorem. Consider a Bell pair of two qubits. One qubit is given to an entity (say Alice) and the other to a different entity (say Bob). At this stage, they each also have access to random variables which share arbitrary correlations. Next, suppose they are given measurements $x$ and $y$ which are not shared, and based on these they produce outcomes $a$ and $b$ respectively. Then quantum mechanics implies that the joint conditional probability distribution, $P(a,b|x,y)$, cannot be reproduced by classical means without communication between Alice and Bob. Further, when $n$ Bell pairs are shared between Alice and Bob, some quantum correlations quantified by $P(a,b|x,y)$, where $x$ and $y$ are selected from a set of $\{0,1\}^{2^n}$ measurements, can only be reproduced by classical means if $\mathcal{O}(2^n)$ number of bits are exchanged between Alice and Bob \cite{Brassard99}.

\begin{figure}
    \centering
    \includegraphics[width=1\columnwidth]{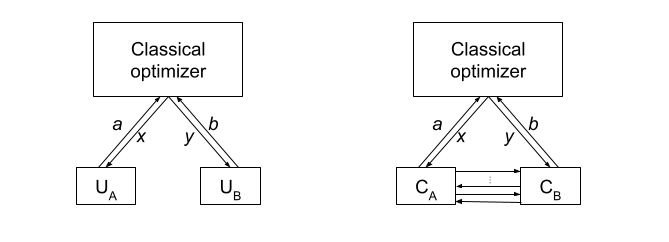}
    \caption{Comparison of quantum (left) vs classical (right) learning. $U_A$ and $U_B$ are unitaries that act on entangled Bell pairs distributed between Alice and Bob. $C_A$ and $C_B$ are operations on random classical bit strings belonging to Alice and Bob.}
    \label{fig:quantum_vs_classical}
\end{figure}

Let us analyze the implication of this for the quantum learning algorithms described in the paper versus classical learning schemes. The communication flow in the quantum learning scheme is shown in the left part of Fig. \ref{fig:quantum_vs_classical}. Once the Bell pairs are created, the classical computer executing the optimization scheme can separately send $x$ and $y$ to Alice and Bob respectively. The measurements $a$ and $b$ are then collectively processed by the classical optimizer to produce the input to the quantum computer for the next iteration.

Now consider how one would simulate the quantum learning algorithm classically. Let us assume that, as is usually the case, the classical algorithm is deterministic in nature, i.e., for every unique input, there is one unique output. In general, the input to the classical generator is a vector of random bits, which has to contain at least $n$ bits for each random variable to produce as many unique outputs as produced in the quantum case. Distribute these bits equally between Alice and Bob. Any classical operation can be written as a combination of operations that only acts on the bits of Alice or Bob separately, and operations that act on both sets of bits. The latter can be further modeled as operations on the bits belonging to Alice, followed by communication to Bob and then operations on just on the bits belonging to Bob, and vice versa (right part of Fig. \ref{fig:quantum_vs_classical}). Then, without loss of generality, we can restrict Alice's operation $C_A$ to be of the form $C_A(x, d_{AB})$, i.e., parametrized by $x$ from the optimizer, and the sequence of bits $d_{AB}$ communicated from Bob to Alice. Similarly, for Bob we have $C_B(y, d_{BA})$. Note that $d_{AB}$ can contain information about $y$ and $d_{BA}$ about $x$, so we have not restricted the form of the classical generator in any fashion. Then, in order to simulate $P(a,b|x,y)$ for arbitrary $x$ and $y$ sent by the classical optimizer, it follows that the amount of communication between Alice and Bob, $d_{BA}+d_{AB}$, will scale as $\mathcal{O}(2^n)$.


\begin{figure}[htb]
    \centering
    \includegraphics[width=1\columnwidth]{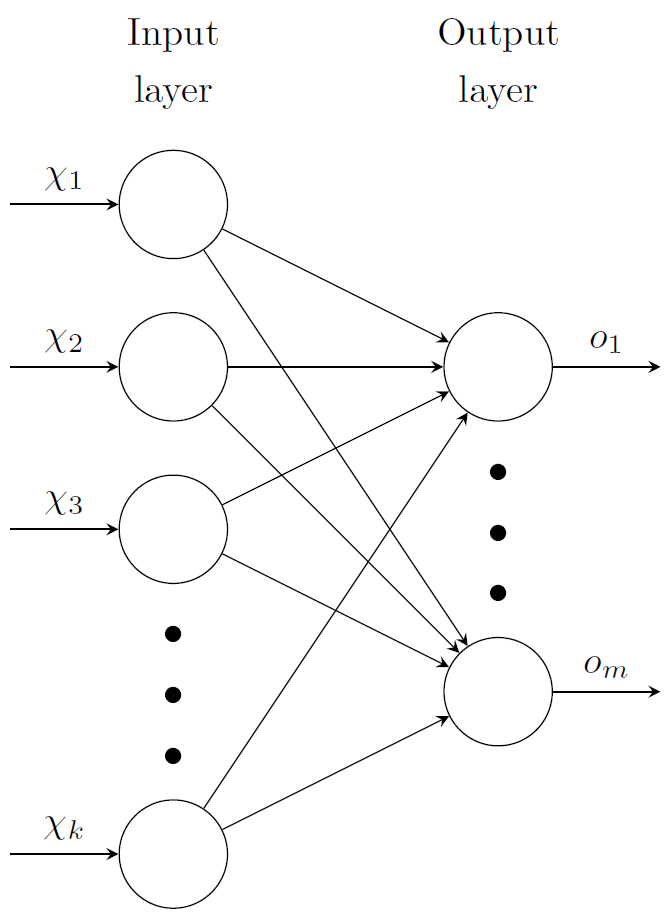}
    \caption{Feed-forward neural network with dimension $[k,m]$.}
    \label{fig:nn}
\end{figure}
Next, let us see the concrete implications of this argument by considering a 2-layer feedforward neural network architecture as shown in Fig \ref{fig:nn}. We assume Alice and Bob share the neurons at each layer and their job is to jointly compute $o=W\chi+b$. Writing it in blocks as
\[
\begin{bmatrix}
o_A\\
o_B
\end{bmatrix}
=
\begin{bmatrix}
W_{A\mathcal{A}}& W_{A\mathcal{B}}\\
W_{B\mathcal{A}}&W_{B\mathcal{B}}
\end{bmatrix}
\begin{bmatrix}
\chi_\mathcal{A}\\
\chi_\mathcal{B}
\end{bmatrix}
+
\begin{bmatrix}
b_A\\
b_B
\end{bmatrix},
\]
where $W_{A\mathcal{A}},W_{B\mathcal{A}},\chi_\mathcal{A},b_A$ belongs to Alice and $W_{A\mathcal{B}},W_{B\mathcal{B}},\chi_\mathcal{B},b_B$ belongs to Bob, Alice locally stores $W_{A\mathcal{A}}\chi_\mathcal{A}$ and sends $W_{B\mathcal{A}}\chi_\mathcal{A}$ to Bob. Similarly, Bob locally stores $W_{B\mathcal{B}}\chi_\mathcal{B}$ and sends $W_{A\mathcal{B}}\chi_\mathcal{B}$ to Alice. After the communication, Alice receives $W_{A\mathcal{B}}\chi_\mathcal{B}$ and computes $o_A=W_{A\mathcal{A}}\chi_\mathcal{A}+W_{A\mathcal{B}}\chi_\mathcal{B}+b_A$, and Bob does so similarly. In this process, Alice sends $\mathcal{O}(|B|)$ bits to Bob and Bob sends $\mathcal{O}(|A|)$ bits to Alice. Since $|A|+|B|=m$, $O(m)$ bits of communication occur.

Thus, if the generator is represented by a deep feed-forward neural network where each layer has at most $m$ neurons, then the number of layers has to scale exponentially asymptotically as $\mathcal{O}(2^n/m)$ to reproduce certain probability distributions generated by experiments on the Bell pairs. This ultimately implies that a neural network will have to have exponentially scaling depth in the precision of the output.

In the argument above, $x$ and $y$ belong to $\{0,1\}^ {2^n}$. For the quantum generator, they can be more compactly specified as angles to the ansatzes specified by $U_A$ and $U_B$ respectively, which can even consist of a constant number of layers. While the full set of measurements takes a circuit of up to depth $\mathcal{O}(2^n)$ to realize, the number of accessible measurements also grows exponentially with depth.

We note that while it is true that qopula circuits exhibit the complexity separations outlined above against classical computation, these are arguments for the worst case and it may not apply to the ones relevant for a particular learning problem.  While determining the generality of these exponential communication and computational separations is a topic of ongoing research, the reasoning above clearly shows that the quantum computer expands the space of joint probability distributions that can be efficiently explored for learning.

\section{Future Work}\label{sec:future}
GANs are prone to mode collapse and nonconvergence, and as such many improvements have been made to mitigate these issues. While our pedagogical case did not suffer from them, they may occur with more complex models and problems. Therefore it would be interesting to see if our framework can benefit from the GAN enhancements discussed in this section. For instance, one may use a critic ($C$) in place of a discriminator. Instead of using binary cross entropy (BCE) as loss function, one may instead use Wasserstein loss (W-Loss) \cite{Arjovsky17}
\begin{equation}
    \min_G\max_C V(G,C)=\mathbb{E}[C(x)]-\mathbb{E}[C(G(z)],
\end{equation}
which tends to make model training more stable. W-Loss solves the vanishing gradient problem of the discriminator, in which the discriminator does not provide enough information for the generator to make progress in the BCE case.

WGAN can be implemented straightforwardly in our case as we can replace the sigmoid function with a linear activation function such that the discriminator becomes the critic and replace BCE with W-Loss as the loss function. We note that this is different from Ref. \cite{Shouvanik19} as we are learning classical data and the discriminator is classical, whereas in Ref. \cite{Shouvanik19} both the discriminator and the generator are quantum. 

Another way to enhance our model is the analog of the addition of noise vectors from the latent space as input to the classical GAN. Recall that in classical GAN, once a model is trained, this input noise vector completely determines the model output (see Sect. \ref{sec:QGANQCBM}). A significant improvement with the addition of input noise vectors is that it allows the controlled generation of data, such as changing specific features of the output by tweaking the noise vector. In our example, this would mean that one can generate samples with desired characteristics such as where all absolute returns are larger than a preset value, as in the case of a black swan event. Controlled generation allows for the emergence of desired features by updating the noise vector according to pre-trained feature classifiers. This would also allow more complex GAN architectures such as styleGAN \cite{Karras19}.

A proposed circuit for such a generator from Sec. \ref{sec:QGANQCBM} allows one to calculate the expectation value of an observable with respect to the noise vector in the latent space,
\begin{align}
    G(z)=\bra{z}U_G^\dag O_G U_G\ket{z}.
\end{align}

$\ket{z}$ may be realized in the following way, where the $Z$ gates are parametrized rotations defined by the noise vector $z$: 
\begin{figure}[htb]
\centering
\hspace{2em}
\Qcircuit @C=1em @R=1.7em {
\lstick{\ket{0}} & \gate{H} & \gate{Z}\\
\lstick{\ket{0}} & \gate{H} & \gate{Z} \\
 &\vdots\\
\lstick{\ket{0}} & \gate{H} &\gate{Z} \\
}
\end{figure}

It is important to note that this model might be more costly to implement on quantum hardware. A circuit execution with 1000 shots corresponds to 1000 data samples in our current model architecture, whereas if using expectation values this would only correspond to one data sample. However, in the long run, such a model might outperform our current model due to the mitigation of problems mentioned above.

Moreover, it is not clear if such a model will be advantageous against classical cases in terms of complexity. While we have argued that producing the distribution in our case is difficult for a classical generator, it might not be the case for producing expectation values as this procedure is deterministic (up to small statistical noise from the measurement). For example, it has been argued that even the lowest depth version of the quantum approximate optimization algorithm (QAOA) can produce distributions, which are hard for classical computers \cite{Farhi16}. However, the computational task of QAOA is to optimize the expectation value $\bra{\boldsymbol{\gamma},\boldsymbol{\beta}}C\ket{\boldsymbol{\gamma},\boldsymbol{\beta}}$ of some objective function $C$ and it is unclear if the low depth version of QAOA can outperform classical algorithms \cite{Hastings18}.

Another area of potential exploration is the development of higher-level ansatzes. Currently, the number of trainable parameters in our ansatz scales quadratically with the number of qubits. This is undesirable as we use more qubits to represent each dimension of data. It may be beneficial to develop ansatz analog to building blocks in classical neural networks (e.g., linear, batch norm), where operations are applied to the variables instead of the bits representing those variables.

Lastly, as discussed in Sec. \ref{sec:QGAN}, instead of utilizing SPSA, the generator can be updated using gradient-based optimizers with the parameter-shift rule. This could allow for more efficient optimization through the cost function. Implementing the parameter-shift rule in parallel would reduce the runtime for quantum gradient estimation. However, this is only possible if the quantum hardware has enough qubits and low crosstalk noise. A networked quantum computer architecture would also make this possible.

\section{Discussion and Conclusions}\label{sec:conclusion}
In our view, the theory of quantum machine learning (QML) has gone through three development phases. The first phase consists of using quantum computers to speed up subroutines of classical learning algorithms \cite{Wiebe12}. These algorithms most often used the quantum linear systems algorithm \cite{Harrow09} as the basis of quantum advantage. The resource requirements for this class of algorithms are very high and it often requires loading classical data in a superposition. The second stage, which started with the advent of cloud-based machine learning, consists of heuristic techniques utilizing variational quantum circuits, which are applied to discriminative or generative problems \cite{Cong2019,ZhuBAS}. While these have been experimentally demonstrated on very small problem sizes, these are heuristic in nature without theoretical support for their advantage over classical methods. Since they cannot be tested at scale, often their efficacy cannot even be numerically predicted. 

The third phase of QML development is now under way. It consists of quantum algorithms with reasonable resource requirements. Moreover, the structure of circuits can often be shown to have advantage over the corresponding classical algorithms. For example, in Ref. \cite{Hsin-Yuan2021}, quantum advantage is proven for certain problems where the objective is achieving a specified worst-case prediction error. Reference \cite{Liu2021} uses the hardness of the discrete log problem to show that a quantum classifier can achieve high accuracy and is robust to additive error where a classical classifier can do no better than random guessing. Perhaps the simplest example of advantage is the proof of separation in expressive power between commonly-used Bayesian networks and their minimal quantum extension shown in Ref. \cite{gao2021enhancing} based on results from quantum foundations. Our paper falls in this third phase. We anticipate that the future of practical QML will follow this path. 

For the particular algorithm presented here, we can project when it will become competitive with classical techniques as follows. Typical structured datasets have dimension of order 10--100. With each dimension represented by 5-10 qubits, we expect our model to have production value if running on quantum hardware of 50 qubits or more.

In conclusion, we have demonstrated quantum generative learning for multivariate data using a QGAN and a QCBM. Our learning models utilize a variational ansatz that we have designed. Our ansatz has provable advantage in capturing correlations between the data over classical methods. We note that our training of the quantum generative models does not show the barren plateau problem and is resilient to noise. The outcome of the QGAN training reaches a better performance and is more stable than training of the equivalent classical GAN.
We anticipate that our observation of a fundamental connection between modeling correlated distributions and quantum entanglement will become the standard technique for modeling multivariate data with quantum computers, and will be used as a basis for a wide variety of learning applications.

\section{Acknowledgement}
This work is a collaboration between Fidelity Center for Applied Technology,
Fidelity Labs, LLC., and IonQ Inc. 

E.Z. and N.M. initiated the idea of doing generative modeling of a copula on quantum computers. E.Z. trained the QGAN, worked on the proof, and worked on the argument for quantum advantage. S.J. designed the qopula circuits, trained the QCBM and worked on the argument for quantum advantage. D.B. worked on the argument for quantum advantage and contributed in all aspects. M.E. contributed to benchmarking the hardware and contributed in all aspects. J.K., J.N., N.P., K.S., K.W. were responsible for the quantum hardware. M.M. and A.S. contributed to scientific discussions. After E.Z. and S.J., all authors are listed in alphabetical order. The Fidelity publishing approval number for this paper is 995073.2.0.
\onecolumngrid
\appendix
\section{Reduced density matrix of a copula state}\label{appd:copulaproof}
Given a quantum state of the form
\begin{align}
    \ket{c}=\sum_{u_1,\cdots,u_d=0}^{m-1}\sum_{k_1,\cdots,k_d=0}^1e^{i\phi(u_1,\cdots,u_d,k_1,\cdots,k_d)}\frac{\sqrt{c(u_1/m,\cdots,u_d/m)}}{2^{d-1}}\ket{u_1,k_1}\cdots\ket{u_d,k_d}.
\end{align}
with $c$ a given discretized copula density, i.e.,
\begin{align}
    \sum_{u_1,\cdots u_{i-1},u_{i+1},u_d=0}^{m-1}c(u_1/m,\ldots,u_d/m)=\frac{1}{m} ~~~\forall i.
\end{align}
Since
\begin{equation}
    \langle v_1,l_1\ket{c}=\sum_{u_2,\cdots,u_d=0}^{m-1}\sum_{k_2,\cdots,k_d=0}^1e^{i\phi(v_1,u_2,\cdots,u_d,l_1,k_2,\cdots,k_d)}\frac{\sqrt{c(v_1/m,\cdots,u_d/m)}}{2^{d-1}}\ket{u_2,k_2}\cdots\ket{u_d,k_d},
\end{equation}
we can see that after tracing over the first register $u_1,k_1$, 
\begin{align}
    \text{Tr}_1(\ket{c}\bra{c})=\sum_{u_1,k_1}&\sum_{u_2,\cdots,u_d}\sum_{k_2,\cdots,k_d}e^{i\phi(u_1,u_2,\cdots,u_d,k_1,k_2,\cdots,k_d)}\frac{\sqrt{c(u_1/m,u_2/m\cdots,u_d/m)}}{2^{d-1}}\ket{u_2,k_2}\cdots\ket{u_d,k_d}\\
    &\sum_{u'_2,\cdots,u'_d}\sum_{k'_2,\cdots,k'_d}e^{-i\phi(u_1,u'_2,\cdots,u'_d,k_1,k'_2,\cdots,k'_d)}\frac{\sqrt{c(u_1/m,u'_2/m,\cdots,u'_d/m)}}{2^{d-1}}\bra{u'_2,k'_2}\cdots\bra{u'_d,k'_d}.
\end{align}
Therefore,
\begin{align}
    \rho_d&=\text{Tr}_{1,\cdots,d-1}(\ket{c}\bra{c})\\
    &=\sum_{u_1,\cdots,u_{d-1},k_1,\cdots,k_{d-1}}\sum_{u_d}\sum_{k_d}e^{i\phi(u_1,\cdots,u_{d-1},u_d,k_1,\cdots,k_{d-1},k_d)}\frac{\sqrt{c(u_1/m,\cdots,u_{d-1}/m,u_d/m)}}{2^{d-1}}\ket{u_d,k_d}\\
    &~~~~~~~~~~~~~~~~~~~\sum_{u'_d}\sum_{k'_d}e^{-i\phi(u_1,\cdots,u_{d-1},u'_d,k_1,k_2,\cdots,k_{d-1},k'_d)}\frac{\sqrt{c(u_1/m,\cdots,u_{d-1}/m,u'_d/m)}}{2^{d-1}}\bra{u'_d,k'_d}.
\end{align}
\twocolumngrid
We see that the requirement of $\rho_d=I/m$ is equivalent to the coefficient of $\ket{u_d,k_d}\bra{u'_d,k'_d}$ being 0 when $u_d\neq u'_d$ or $k_d\neq k'_d$. This can always be achieved by setting the phases $\phi(u_1,\cdots,u_d,k_1,\cdots,k_d)$ appropriately.

In total, by enforcing $\rho_i=I/m$ for all $i$, we will end up with $dm(2m-1)$ number of equations. However, the number of degrees of freedom that we have in setting the phases is $(2m)^d-1$. Since $(2m)^d-1>dm(2m-1)$ whenever $d\geq 2$ and $m\geq 2$, we will always have enough degrees of freedom to ensure maximal entanglement in $\ket{c}$.

In practice, the extra qubits $\ket{k_1},\cdots,\ket{k_d}$ are usually not needed unless the copula function $c$ has extreme spikes in certain regions.

\section{QCBM Training}\label{appd:QCBMtraining}
\begin{figure}[h]
\centering
        \includegraphics[width=1\columnwidth]{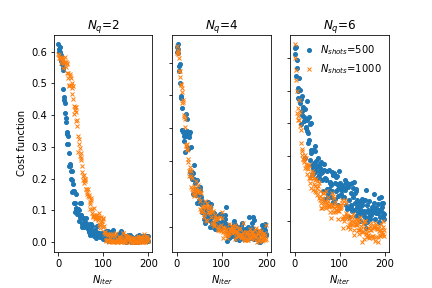}
    \caption{Effect of increasing shots on the convergence of the QCBM for different system sizes.}
    \label{fig:shots_qcbm}
\end{figure}

Here we discuss details of the QCBM training. Figure \ref{fig:shots_qcbm} shows the effect of changing the number of shots $N_{shots}$ on the simulator. For $N_q=2$, the cost function converges faster for fewer shots while this is reversed for $N_q=6$. This further supports our observation in the text that for the QCBM at small system sizes, a noisy estimate of the cost function can play a favorable role in the optimization. 
For $N_q=6$, one can also observe that the spread of the values towards the end of the optimization decreases due to the increased accuracy available from larger number of shots. 

\begin{figure}[h]
\centering
        \includegraphics[width=1\columnwidth]{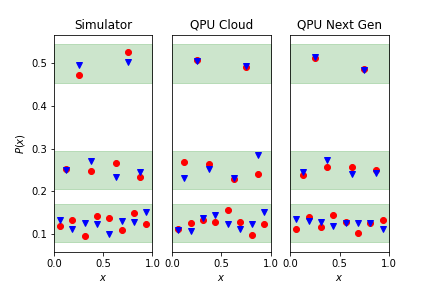}
    \caption{Marginal distributions in the copula space on the three different backends used. The red circles correspond to the first variable and the blue triangles correspond to the second. The width of the green-shaded bands is the theoretical statistical uncertainty, which is $1/\sqrt{N_{shots}}$ and they are centered at $1/2^{N_q/2}$, where $N_q=$ 2, 4, and 6 qubits from top to bottom in each subfigure. Here $N_{shots}=500$.}
    \label{fig:copula_qcbm}
\end{figure}

Figure \ref{fig:copula_qcbm} shows the marginal distributions obtained at the minimum value of the cost function achieved during optimization. Theoretically, the marginals should be uniform within statistical error. This is seen to be true even on the quantum hardware, revealing one reason for the robustness of the algorithm to noise.

\section{Proof of quantum advantage based on Bell's Theorem}\label{appd:bell}
\begin{theorem}
\textit{Simulating Algorithm \ref{alg:QGAN} or Algorithm \ref{alg:QCBM} on a classical deterministic computer using the qopula ansatz with $n$ qubits per variable will take an amount of time $O(\exp(n))$}.
\end{theorem} 

In each iteration of the Algorithms \ref{alg:QGAN} and \ref{alg:QCBM}, the first step is to create $n$ Bell pairs and distribute them between two registers A and B. Let us say the classical computer executing the optimization scheme sends parameter vectors $x$ and $y$ to the quantum computer. The unitary $U_A$ acting on register A is a function of $x$ and the unitary $U_B$ is a function of $y$. From the point of view of the quantum computer, $x$ and $y$ are arbitrary. After execution of the qopula circuit, the measurements $a$ and $b$ from A and B respectively are then collectively processed by the classical optimizer to produce the input to the quantum computer for the next iteration. The measurements are then samples from a joint probability distribution, say $P(a,b|x,y)$. 

Now consider how one would simulate these quantum learning algorithms classically. The classical algorithm is deterministic in nature, i.e., for every unique input, there is one unique output. In general, the input to the classical generator function is a vector of random bits, which has to contain at least $n$ bits for each random variable to produce as many unique outputs as produced in the quantum case. Distribute these bits equally between two registers A and B. Any classical operation can be written as a combination of operations that only acts on the bits of A or B separately, and operations that act on both sets of bits. The latter can be further modeled as operations on the bits belonging to A, followed by communication to B and then operations on just on the bits belonging to B, and vice versa (right part of Fig. \ref{fig:quantum_vs_classical}). Then, without loss of generality, we can restrict $C_A$ to be of the form $C_A(x, d_{AB})$, i.e., parametrized by $x$ from the optimizer, and the sequence of bits $d_{AB}$ communicated from B to A. Similarly, for B we have $C_B(y, d_{BA})$. Note that $d_{AB}$ can contain information about $y$ and $d_{BA}$ about $x$, so we have not restricted the form of the classical generator function in any fashion. Then, in order to simulate $P(a,b|x,y)$ for arbitrary $x$ and $y$ sent by the classical optimizer, it follows Theorem 4 of Ref. \cite{Brassard99} that the amount of communication between A and B, $d_{BA}+d_{AB}$, will scale as $\mathcal{O}(2^n)$. Thus, the computation of $C_A$ and $C_B$, which is required to simulate Algorithms 1 and 2 will also take time $\mathcal{O}(2^n)$.
\appendix



\begin{thebibliography}{10}
\bibitem{ph}The polynomial hierarchy class PH \cite{Arora09} is defined to be the union of an infinite tower of increasing classes $\Delta_k$,$k=1,2,...$, in which $\Delta_1=P$ and $\Delta_{k+1}=P^{N\Delta_k}$.\ignorespaces
\end{thebibliography}
\end{document}